\documentclass[12pt,preprint]{aastex}

\newcommand\lam{\mbox{$\:\lambda $ }}

\newcommand\kms{\:\rm{\,km\,s^{-1}}}

\newcommand\peryr{\:\rm{\,yr^{-1}}}

\newcommand\perpix{\:{\rm pixel}^{-1}}

\newcommand\chandra{{\it Chandra}}

\newcommand\hi{\ion{H}{1}}

\newcommand\oiii{[\ion{O}{3}]}

\slugcomment{Accepted for publication in  {\it Ap.\ J.} }

\shorttitle{Neutron Star Recoil in Puppis A}
\shortauthors{Winkler \& Petre}

\begin{document}


\title{Direct Measurement of Neutron-Star Recoil in the Oxygen-Rich Supernova Remnant Puppis A}


\author{P. Frank Winkler }  
\affil{Department of Physics, Middlebury College, Middlebury, VT 05753}
\email{winkler@middlebury.edu}
\and
\author{Robert Petre } 
\affil{NASA Goddard Space Flight Center, Greenbelt, MD}
\email{robert.petre-1@nasa.gov}



\begin{abstract}

A sequence of three {\it Chandra} X-ray Observatory High Resolution Camera images taken over a span of five years reveals  arc-second-scale displacement of RX~J0822--4300, 
the stellar remnant (presumably a neutron star) near the 
center of the Puppis A supernova remnant.  
We measure its proper motion to be $0.165 \pm 0.025$\ arcsec yr$^{-1}$\ toward the west-southwest.
At a distance of 2 kpc, this corresponds to a transverse space velocity of $\sim$1600 km s$^{-1}$.  
The space velocity is consistent with the explosion center inferred from proper motions of the
oxygen-rich optical filaments, and confirms the idea that Puppis A resulted from an asymmetric 
explosion accompanied by a kick that imparted roughly 3$\times$10$^{49}$ ergs 
of kinetic energy (some 3 percent of the kinetic energy for a typical supernova) to the stellar remnant.  
We discuss constraints on core-collapse supernova models that have been proposed to explain 
neutron star kick velocities.  

\end{abstract}


\keywords{ISM: individual (Puppis A) --- X-rays: individual (RX~J0822--4300) --- stars: neutron --- 
(stars:) supernovae: general ---  (ISM:) supernova remnants --- astrometry}


\section{Introduction}

There has long been broad consensus that core-collapse supernovae---the 
explosions of massive progenitors that produce Types II, Ib, and Ic 
events at least---leave behind 
a compact stellar remnant: either a neutron star or a black hole.  
Early on, this model was marred by the paucity of observed compact 
objects associated with supernova remnants (SNRs).  The  
discovery in recent years of numerous compact X-ray sources associated with SNRs, especially 
with the oxygen-rich SNRs that are presumably the young remnants of 
core-collapse SNe, has removed this blemish \citep[e.g.][for a recent review]{manchester01}. 
Compact stellar remnants have been identified near the centers of all three of the known oxygen-rich SNRs in the Galaxy---Cas A \citep{tananbaum99, chakrabarty01}, Puppis A \citep{petre96}, and most recently 
G292.0+1.8 \citep{hughes01, camilo02}---plus 0540--69.3 in the LMC \citep{seward84}.

There is also growing evidence, both observational and theoretical, that 
the explosions of core-collapse SNe are asymmetric.
In young, oxygen-rich SNRs, whose fast filaments provide the best 
opportunities for observing ejecta from the cores of massive SNe, 
anisotropies appear to be typical:  e.g., Cas A  \citep{lawrence95, reed95, fesen01, hwang00, hwang04}, 
and the LMC remnants N132D  \citep{morse95}, and 0540--69.3 \citep{kirshner89}, 
in addition to Puppis A \citep{winkler85}.  Furthermore, 
two-dimensional models for core-collapse SNe show instabilities just prior 
to collapse that produce asymmetries in the ejecta distribution \citep[e.g.,][]{burrows95}.

Simple conservation of momentum requires that if the explosion of a 
progenitor star expels the ejecta preferentially in one direction, the 
compact core must recoil in the opposite direction.  By measuring 
the displacements of young pulsars from the apparent centers of 
their associated SNR shells and using the pulsar spin-down periods as 
age estimates, \citet{caraveo93} and \citet{frail94}
inferred that pulsars are typically born with transverse velocities of 
$500 \kms$, and that velocities $\gtrsim 2000 \kms$\ may occur.
Direct radio and optical measurements have found transverse velocities ranging from 
$60\kms $\  \citep[for Vela -- ][]{dodson03} to $240 \kms$\  
\citep[for PSR B1951+32 in CTB~80 --][]{migliazzo02}.

Here we report measurement of the recoil of RX~J0822--4300, the presumed neutron star near
the center of Puppis A, obtained through images taken at three epochs spanning 5 years
from the High Resolution Camera (HRC) on the {\it Chandra} X-ray Observatory.
Past optical studies of the most prominent ejecta-dominated filaments 
show that they are concentrated in the northeast quadrant of the remnant, and that 
their motions are northward and eastward, consistent with undecelerated 
expansion from a common center \citep{winkler85, winkler88}.   The present
measurement of the stellar remnant's rapid motion to the west-southwest completes a 
picture of asymmetric ejecta and neutron star recoil resulting from a core-collapse supernova.

As the original version of this  paper was nearing completion, we became aware of a paper
by \citet[][henceforth HB06]{hui06a} that presents a similar analysis based on two of the three {\it Chandra} 
observations used here.  While the HB06 result is qualitatively similar to our own, our
analysis leads to a more precise measurement and a significantly higher velocity for 
RX~J0822--4300.  We compare their methods with our own in \S 3.5.

\section{Observations}
The compact X-ray source near the center of Puppis A, RX~J0822--4300, was observed
with the HRC-I in December 1999 
(OBSID 749), 
and again with the HRC-S in January 2001 (OBSID 1851).  
We repeated the earlier of these previous observations, 
using the HRC-I, on 25 April, 2005 (OBSID 4612).   
For all three observations,  RX~J0822--4300 was 
placed essentially on-axis for optimum spatial resolution.  The observational details are 
summarized in Table 1.

At the $\sim 2$\ kpc distance of Puppis A, a transverse velocity of $1000\kms$\ would give 
a proper motion of only $\sim 0\farcs1 \:{\rm yr}^{-1}$.  Given {\it Chandra's} uncertainty of 
$\sim 0\farcs 5$\ in absolute aspect, measurement of the motion over a baseline of
a few years is a challenge.   By great good luck,
there are two additional point sources quite close to RX~J0822--4300, almost optimally situated
for a proper motion study.   The closer, which we refer to as Star A,  is located 2\farcm7  to the 
southwest of the stellar remnant; 
the other, Star B, is 5\farcm4 to the northeast.  
Both coincide with  
$V \sim 13$--14 mag stars included in the UCAC2 astrometric catalog 
\citep{zacharias03}, with  precisely measured (within 15--24 mas) positions and 
proper motions.   Therefore, it is possible to use these stars as fiducial sources to 
provide a precise absolute coordinate system for the image at each epoch.

For each of the three HRC images, we first identified the sources and measured their 
positions using the standard WAVDETECT routine in the Chandra Interactive 
Analysis of Observations  (CIAO) software package (version 3.3).  
RX~J0822--4300  and both fiducial stars were detected with 
high signal-to-noise ($> 5\sigma$) in all three observations, with positions 
within 0\farcs6 of the UCAC2 catalog positions.
There can be no doubt that the association of the X-ray sources 
with the UCAC2 stars is correct.  
In addition, there is another, very faint, X-ray source in the field at a position only 
2\farcm0 from RX~J0822--4300 and coincident within 0\farcs5 with a third UCAC2 star (Star C).  
This source was detected at the $\sim 3\sigma$\ level in the first two \chandra\ observations (1999 and 2001), but the detection was extremely marginal ($\sim 5\ {\rm counts}\lesssim 2\sigma$) in the
2005 observation.
In order to maintain consistency, we did not use it for the astrometry in any of the images.

Relevant data for all three stars, extracted from the UCAC2 catalog, appear in Table 2.   
All three also appear in the 2MASS catalog \citep{cutri03} at positions consistent with those in 
the UCAC2 but with lower precision.  
In Figs.\ 1 and 2, respectively, we show the 
central region of an image from the 2005 HRC-I observation with the sources marked, 
and the identical field in an optical image  taken with the CTIO 0.9 m telescope
through a  narrow-band red filter (CWL 6852 \AA, FWHM 95 \AA).

\section{Precise Astrometry and Proper Motion Measurement}
As a preface to this section, we note that the proper motion of RX~J0822--4300 (which we
refer to more succinctly as the neutron star, NS) is apparent in the data.   
Simple comparison between images based on the original level-2 event files 
with the nominal aspect
shows a displacement of order 1\arcsec\  west, and slightly south, from the
Dec 1999 HRC-I observation to the one in April 2005.  
The difference  is noticeable by blinking the two images,
and is even more apparent when one scales the images by the exposure time and 
subtracts one from the next.   

The absolute aspect of {\it Chandra} HRC images may be subject to uncertainty 
of order 0\farcs 5, however, so we have gone to considerable effort to achieve
the highest accuracy possible in the astrometry, based on the fiducial stars A and B\@.
We discuss below two techniques, the first a simple translation from one epoch
to the other (applicable for the HRC-I images only), and the second involving transformation 
of all three images to an absolute world coordinate system based on the optical positions for the 
fiducial stars.   Both techniques begin with precise measurement of the X-ray
positions of the NS and stars A and B on each HRC image, discussed in \S3.1.  The subsequent
sections describe the two transformation techniques and proper-motion measurements.
Since the measurement involves pushing HRC astrometry to its limits, we describe 
these steps in some detail.  

\subsection{Nominal X-ray Positions}
For the final analysis we reprocessed the {\it Chandra} level-1 event files to incorporate
the latest degapping and tap-ringing corrections.  We then 
used the \chandra\ ray-tracing and simulation (ChaRT/MARX) routines to 
generate off-axis point-spread functions (PSFs) appropriate for each of the three
sources at each epoch, following the ChaRT threads ``Using MARX to Create an Event File" and
``Creating an Image of the PSF,"
and normalized these to match 
the observed counts in each image.    
We then followed the CIAO/Sherpa thread ``Using a PSF Image as the Convolution Kernel"
to obtain the best-fit position  for each source.   This procedure convolves the appropriate PSF
with a source model, and varies the model parameters to achieve the best match to the data. 
For the NS and star A we used the 
original (unbinned) data with a scale of $0\farcs 1318\; \perpix$, but for star B we binned the data
(and the matching PSF) $2 \times 2$.

As a source model we used a narrow Gaussian with FWHM fixed at 1 pixel 
$\simeq 0\farcs 13\ (\ll$ the PSF width, even for a near on-axis source),\footnote{The alternative of a model using a true 
point source leads to similar results for the fitted position, but (as currently implemented in Sherpa) fails to produce maps of confidence contours with sufficient resolution to give precise uncertainties.  An example of such a map for the narrow Gaussian model we actually used is in panel ({\it d)} of Fig.~3.} and varied four parameters:  
$x$\ and $y$\ position (RA and Dec, respectively), source strength, and (constant) background level.    
Convolving this model with the appropriate PSF matches the original data well in all cases.  
We also experimented with varying the width of the Gaussian, and found that the fits are insensitive
to the choice of FWHM as long as it is $\lesssim 3$\ pixels.   There is no evidence for a finite 
extent to the NS source (nor for either of the stellar sources).  In Fig.~3 we show an example of
one of the position fits, for star A at the 2005 epoch. 

The results of the position fits for all three sources
(based on the nominal aspect solution) at all three epochs are given in Table 3.   All are
based on fits like the one illustrated in Fig.~3.  The uncertainties
represent the 1-$\sigma$\ (68\% confidence) limits for the source position along the $x$\ and $y$\ axes.  
To obtain these, we used Sherpa to calculate the \citet{cash79} statistic over a grid and 
used an increment of 2.3 above the minimum as the estimate of the 1-$\sigma$\  confidence limit for a model with two ``interesting"
parameters \citep{press92}.  An example of a Cash-statistic contour plot resulting from this 
procedure is shown in Fig.~3, panel ({\it d}).  We note  that the contours are not circular; 
when projecting along
a different axis (as we do in \S 3.4) we used the actual values appropriate for that direction, as measured
graphically on these plots. 
 
\subsection{Method 1: Simple Translations}
The simplest approach to measuring the NS proper motion is a two-dimensional 
translation of the image from the second
epoch relative to the first, based on the fiducial stars.   If all aspects of the instrument and the
data train are identical, this ought to bring the fiducial stars into alignment.   
But translation alone is only modestly successful in 
aligning stars A and B in the pair of images from the two HRC-I epochs.
Even after correcting for the (small) proper motions of the stars (taken from the UCAC2) between 
epochs, we find shifts for stars A and B that differ by 
$(0\farcs95 \pm 0\farcs31, -0\farcs71 \pm 0\farcs42)$\ in $x$\ and $y$, respectively.  
An average shift weighted inversely as the variance is dominated by the shift for  
the more nearly on-axis star A, and leads to a change in the NS position of 
$(- 0\farcs 71\pm 0\farcs 12, -0\farcs 25\pm 0.20)$\ in RA and Dec, respectively, from epoch
1998.98 to 2005.32, or proper motion $0\farcs 141 \pm 0\farcs 024\peryr$\ at 
an angle of $20\degr \pm 14\degr$\ south of due west.  The uncertainties are purely formal, based on 
uncertainties in the position fits for stars A and B, the (much smaller) uncertainties in
the proper motions for the optical stars, and the (negligible) uncertainties in the position
fit for the NS at the two epochs (all as given in Table 3).    We did not attempt a similar 
pure translation between the HRC-S and HRC-I images, since there is a known small 
rotational offset between the HRC-S and HRC-I, and possibly other small systematic
effects as well (R. Kraft, private communication).  

\subsection{Method 2:  Transformation to an Absolute Frame}
A more sophisticated technique follows precise measurement of the X-ray
positions of the NS and Stars A and B on each HRC image (\S3.1) with
transformation of both HRC images
to a common {\it absolute} world coordinate system (WCS) based on the cataloged 
optical positions for stars A and B\@.   Then the NS position at each epoch, its motion
between epochs, and uncertainties in these quantities may be  determined 
from the transformations.  
We assume that the ``true" positions for stars A and B are those for the optical stars given
by the UCAC2, corrected for proper motion to the appropriate epoch.   These are also given 
in Table 3.    We then calculate the transformation so that the measured X-ray positions 
will exactly match the optical ones at each epoch.  

We assume a linear transformation that  involves 
four parameters:  translations 
in $x$\ and $y$\ ($t_x$\ and $t_x$), rotation through an angle $\theta$, 
and a uniform change in scale by a factor $r$\@.   With only two fiducial 
points it is straightforward to calculate a unique transformation analytically.
The transformation parameters are determined 
from $(x_A, y_A),\, (x_B, y_B)$, the $x$\ and $y$ positions for stars A and B as 
measured in an HRC image, and  $(x'_A, y'_A),\, (x'_B, y'_B)$, the reference positions 
from the UCAC2 at the same epoch.   The transformation is calculated by
\begin{equation}
\left( {\begin{array}{*{20}c}
   { - y_A } & {x_A } & 1 & 0  \\
   {x_A } & {y_A } & 0 & 1  \\
   { - y_B } & {x_B } & 1 & 0  \\
   {x_B } & {y_B } & 0 & 1  \\
\end{array}} \right)\left( {\begin{array}{*{20}c}
   p  \\
   q  \\
   {t_x }  \\
   {t_y }  \\
\end{array}} \right) = \left( {\begin{array}{*{20}c}
   {x'_A }  \\
   {y'_A }  \\
   {x'_B }  \\
   {y'_B }  \\
\end{array}} \right),
\end{equation}
where $p = r\:{\rm sin}\theta$\ and $q = r\:{\rm cos}\theta$.  Inverting the matrix leads to the transformation parameters $t_x,\, t_y,\, r,\, \theta$,  which we then apply to the
position of the NS measured on the \chandra\  image  to find its
corrected position at that epoch,
\begin{equation}
\left( {\begin{array}{*{20}c}
   {x'_{NS} }  \\
   {y'_{NS} }  \\
\end{array}} \right) = \left( {\begin{array}{*{20}c}
   q & { - p}  \\
   p & q  \\
\end{array}} \right)\left( {\begin{array}{*{20}c}
   {x_{NS} }  \\
   {y_{NS} }  \\
\end{array}} \right) + \left( {\begin{array}{*{20}c}
   {t_x }  \\
   {t_y }  \\
\end{array}} \right).
\end{equation}
  
The results of  our transformation analysis indicate corrections to the nominal aspect that
are small, but significant at the sub-arcsecond level.  The stretch factors $r$\ required to scale
from the original frame to the absolute one  are 1.0022(6), 1.0033(3), and 1.0002(2) 
for data from the 1999 HRC-I, 2001 HRC-S, and 2005 HRC-I observations, respectively,
where the number in parentheses represents the uncertainty in the final digit.  
The rotation angles (in the same order) are $-0\fdg01(4), +0\fdg 12(2)$, and 0.00(1).  
The translations in $x$\ and $y$\ are all $\lesssim 0\farcs 5$. 

The adjusted position for the NS at each epoch and its proper motion between epochs 
are given in Table 4.   
The two pairs of measurements with baselines longer than 4 years--- 
OBSIDs 749 vs.\ 4612 (5.34 yr), OBSIDs 1851 vs.\ 4612 (4.25 yr)---both 
yield results that are very similar to one another and to the measurement translation-only 
analysis for the 749-4612 pair (\S3.2): 
a displacement corresponding to a proper motion of $\sim 0\farcs 16 \peryr$\ 
in a direction 15\degr--30\degr\ south of west.  The total displacement from 1999 December until 2005 April is 0\farcs 75 based on pure translation, or 0\farcs 99 based on the more general transformation.
From 2001 January until 2005 April the displacement is 0\farcs 72.  
For  our overall measurement we adopt the unweighted mean of all three measurements:
$0\farcs 165 \pm 0\farcs 025\; {\rm yr}^{-1}$\ at  position angle $248\degr \pm \, 14\degr$.
Whatever the technique used, the statistical uncertainty is dominated by uncertainty in 
measuring the X-ray positions for the reference stars.   We discuss this further in \S3.4.

In Fig.\ 4 we show a difference image between the December 1999 and April 2005 exposures 
(registered and scaled by the exposure time).  The only strong feature is at the position of the stellar remnant.  The inset shows a small region surrounding the stellar remnant from Figs.\ 1 and 4.  The evidence for movement of RX~J0822-4300 is unequivocal.


\subsection{Uncertainty Estimate, and a Simplification}

The variance in $x'_{NS}$\ and $y'_{NS}$\ may be calculated by 
combining in quadrature the contributions from each of the 10 measured quantities,
$x_A,\, y_A,\, x_B,\, y_B,\, x'_A,\, y'_A,\, x'_B,\, y'_B,\, x_{NS}$, and $y_{NS}$, by
\begin{equation}
\sigma _{x'_{NS} }^2  = \left( {\frac{{\partial x'_{NS} }}{{\partial x_A }}} \right)^2 \sigma _{x_A }^2  + \left( {\frac{{\partial x'_{NS} }}{{\partial y_A }}} \right)^2 \sigma _{y_A }^2  +  \cdots  + \left( {\frac{{\partial x'_{NS} }}{{\partial y_{NS} }}} \right)^2 \sigma _{y_{NS} }^2 .
\end{equation}
For finding the {\it magnitude} of the proper motion, this  tedious calculation can be considerably simplified by a very fortuitous accident:
the NS position lies not far off the line joining stars A and B, and furthermore 
the direction of its motion is nearly parallel to this line.    We can thus work in a coordinate
system $(u, v)$\ that is rotated by $30\fdg5$\   clockwise with respect to the $(x, y)$\ system, so that
the $u$\ axis runs parallel to the line from B to A\@.\footnote{While this line will be in slightly different directions at the different epochs, due to 
proper motions of A and B, we have used the orientation at the 2005  epoch for all.  Proper motions
of the reference stars lead to slightly different $(u,v)$\ coordinates for each at the 
different epochs, but the contribution of differences in $v$\ to the scale factor are negligible.}
The original two-dimensional problem may then be approximated by a one-dimensional one:
finding the position of an intermediate point along an elastic band that we allow to stretch 
uniformly between end points at known locations.  
Based on only five measured quantities---the $u$\ coordinates for the three X-ray sources, which we denote as $a$, $b$, and $n$ for stars A, B, and the NS, respectively, and 
the corresponding optical coordinates for stars A and B, denoted as $a^\prime$\ and $b'$---we can calculate
$n'$, the true position for the NS along the $u$\ axis, along with its uncertainty $\sigma_{n'}$.

In this one-dimensional approximation, the transformation from the measured X-ray position $u$ 
at any epoch to the reference (optical) position is given simply by $u' = ru + t$, where $r$\ is a stretch 
factor and $t$\ a translation.  
Writing the transformation for the reference points $a$ and $b$, we immediately find
 \begin{equation}
r = \frac{{b' - a'}}{{b - a}}, \ \ \ \ t = a' - \frac{{b' - a'}}{{b - a}}a,
 \end{equation}
and thus $n'$\ is
 \begin{equation}
n' = a' + \frac{{b' - a'}}{{b - a}}\left( {n - a} \right).
 \end{equation}
 
 In the usual manner for obtaining the variance in a quantity that depends on several independent variables.  
 \begin{equation}
\sigma _{n'} ^2  = \left( {\frac{{\partial n'}}{{\partial a}}} \right)^2 \sigma _a ^2  + \left( {\frac{{\partial n'}}{{\partial b}}} \right)^2 \sigma _b ^2  + \left( {\frac{{\partial n'}}{{\partial n}}} \right)^2 \sigma _n ^2  + \left( {\frac{{\partial n'}}{{\partial a'}}} \right)^2 \sigma _{a'} ^2  + \left( {\frac{{\partial n'}}{{\partial b'}}} \right)^2 \sigma _{b'} ^2 .
\end{equation}
Straightforward calculation of the partial derivatives gives results such as 
 \begin{equation}
\frac{{\partial n'}}{{\partial a}} = \frac{{\left( {n - b} \right)\left( {b' - a'} \right)}}{{\left( {b - a} \right)^2 }} = r\frac{{\left( {n - b} \right)}}{{\left( {b - a} \right)}} \approx \frac{{\left( {n - b} \right)}}{{\left( {b - a} \right)}},
\end{equation}
 where the final approximation utilizes the fact that the stretch factor $r$\ is very nearly 1.
With similar approximations for the other derivatives, (7) becomes
 \begin{equation}
\sigma _{n'} ^2  \approx  \left( {\frac{{n - b}}{{b - a}}} \right)^2 \left( {\sigma _a ^2  + \sigma _{a'} ^2 } \right) + \left( {\frac{{n - a}}{{b - a}}} \right)^2 \left( {\sigma _b ^2  + \sigma _{b'} ^2 } \right) + \sigma _n ^2. 
 \end{equation}
 
 The individual uncertainties in Table 4 are calculated in this way and include 
 the formal uncertainties in fits to the X-ray positions for stars A and B and for 
RX~J0822--4300, and also the (smaller) 
position uncertainties from the UCAC2 catalog, for both epochs.  
We have included proper-motion uncertainties 
from the UCAC2, so the overall position uncertainty for the astrometric stars increases with time 
since the 2000.0 reference epoch.   However, uncertainties in the X-ray source positions
still dominate at all three epochs.     Not surprisingly, the statistical uncertainties in 
all three proper-motion determinations are comparable.   
For our final result, we have taken a conservative approach:  an unweighted average 
of the three measurements, 
with an uncertainty comparable to that of any of the individual ones and large enough to 
embrace them all.

\subsection{Comparison with Hui \& Becker (2006a)}

In their  analysis,  HB06 have obtained a qualitatively similar but
smaller and more uncertain value for the proper motion:  
$0\farcs 104 \pm 0.040\; {\rm yr}^{-1}$\ at  position angle $240\degr \pm 28\degr$.
Their analysis differs from our own in several respects:  
(1) They used only the two HRC-I
observations, while we have also included the 2001 observation with the HRC-S. 
(2) HB06 used only the closer of the two stars in the field (star A in our nomenclature; 
star B in theirs) as an astrometric reference, while we used both.
(3) HB06 based their position fits
on PSF's interpolated from the library in CALDB, 
while ours are based on PSF's we generated specifically for each source and observation 
using ChaRT and MARX. (The CIAO thread ``Why use ChaRT instead of the PSF libraries?"  
explicitly discusses this difference.)
(4) HB06 took the absolute position of their reference star from the  
2MASS catalog, whereas we used the somewhat more precise ones from UCAC2 that also include
proper-motion corrections.   
Despite these differences, the fact that independent analyses carried out by different groups
lead to results that are {\it qualitatively} similar---the neutron star recoiling to the west-southwest
with  high  velocity---further supports the robustness of the result.  However, there are significant
quantitative differences:  HB06a found $\mu = 0\farcs 104 \pm 0.040\; {\rm yr}^{-1}$\ vs. 
our measurement of $0\farcs 165 \pm 0.025\; {\rm yr}^{-1}$\@.   We discuss some implications of
such a high proper motion and the implied transverse velocity  in the next two sections.

\section{Space Velocity and Kinematics}

The most recent estimate for the distance to Puppis A is by \citet{reynoso95}, who found
$2.2 \pm 0.3$ kpc, based on the velocity of \hi\ absorption features.  An independent estimate is
$1.8 \pm 0.5 $ kpc, based on a possible association with Vela OB1 \citep{sakhibov83}.  For 
simplicity, we calculate the space velocity scaled to a distance $d_2 \equiv d/2\: {\rm kpc}$\@.   
The measured proper motion of $0\farcs 165 \pm 0.025\; {\rm yr}^{-1}$\ corresponds to a transverse space velocity of $1570\pm 240\,d_2 \kms$\@. 
The high proper motion confirms the general picture of an asymmetric explosion of the Puppis A progenitor, accompanied by a violent kick to the stellar remnant; a neutron star space velocity exceeding $\sim$100 km s$^{-1}$ cannot be produced by the break up of a binary system by a supernova.

\citet{petre96} predicted just this sort of motion for RX~J0822-4300 based on its position 
6\arcmin\ west-southwest of the expansion center for the oxygen knots measured by \citet{winkler88}.  
For an age of 3700 yr, also determined by the knot kinematics, the expected transverse velocity
is $980\,d_2 \kms$.   As shown in Fig.\ 5, extrapolation backwards from RX~J0822-4300
along the vector representing the measured proper motion almost grazes the 
90\% confidence contour for the explosion center.  The present measurement 
provides strong qualitative confirmation for the picture of a recoiling neutron star as
laid out by \citet{petre96}, but it appears that the actual velocity is $\sim$50\% higher than
they predicted.   This suggests that the Puppis A remnant is somewhat younger than 3700 yr, and/or
that the true expansion center is somewhat west and south of the one found by \citet{winkler88}.  

How asymmetric was the explosion?   
Given a velocity of $1570\, d_2 \kms$\ and a nominal neutron star mass of 1.4~M$_{\odot}$, the kinetic energy of associated with the compact star is $\sim 3\times10^{49}\,d_2^2\:$ ergs.  This represents 3\% of the total kinetic energy of 10$^{51}$ ergs produced by a canonical supernova explosion. Conservation of momentum requires that the net momentum of material ejected in the 
opposite direction is  
$1.4 M_{\sun}\times1570 \kms \sim 4\times10^{41}$ g cm s$^{-1}$.  

Overall the Puppis A remnant looks reasonably symmetric in X-rays after the northeasterly gradient in the density of the ambient medium is taken into account.  Any asymmetries associated with the forward shock have long been submerged by asymmetries in the ISM.  Oxygen is the most prominent ejecta species: the {\it Einstein} FPCS detected a substantial overabundance of highly ionized oxygen at  $\sim 2\times 10^6$\ K.  Assuming the oxygen is uniformly dispersed throughout the remnant, \citet{canizares81} estimated a total oxygen mass of $>$3~M$_{\odot}$, from which they inferred a progenitor mass of $>$25~M$_{\odot}$.  Subsequent X-ray spectral imaging has not revealed any asymmetry in oxygen or any other ejecta species \citep{tamura94}.  The only manifestation of asymmetry is the array of fast moving knots, composed almost entirely of warm ([O~III]-emitting) oxygen and neon \citep{winkler85, winkler88}.

It is possible to estimate whether sufficient mass is contained in these knots to balance the momentum.  
\citet{winkler88} give 0.1 pc and 200 O atoms cm$^{-3}$\  as the typical knot size and density, for a mass $\sim 0.04 M_{\sun}$.  The proper motions for the 11 measured knots 
correspond to 1,000-2,500 km s$^{-1}$\ at 2 kpc.  
For a typical velocity component opposite the direction of the stellar remnant of 1,500 km s$^{-1}$, the corresponding momentum per knot 
is $\sim1.2\times10^{40}$ g cm s$^{-1}$.  Thus $\sim 30$\ such knots are needed to offset the momentum of RX~J0822--4300\@.  
Fig. 5 suggests the existence of this number of knots is reasonable.  

\section{Discussion}

The nature of RX~J0822--4300 is not well understood.  Like most other objects lumped into the ``central compact object" class, it is detected only in X-rays and shows no strong long- or short-term temporal variability.  \citet{hui06b} find a candidate period of $\sim$0.22 s, with a 5$\pm$1 percent pulsed fraction.\footnote{The 75 ms period  once-proposed for RX~J0822--4300  has not withstood further scrutiny  \citep{pavlov02,hui06b}.}  Its X-ray spectrum and flux are consistent with thermal emission from a neutron star surface.  \citet{pavlov02} find acceptable fits using either a black body with 
$kT \sim 0.4\,$ keV and an emitting radius $R \sim1.4\:$ km, or a hydrogen atmosphere model with 
$kT_{eff}^{\infty} \sim0.17\,$ keV, an emitting radius $R_{\infty}\sim 10\,$ km, and surface magnetic field strength 
$B \gtrsim 6 \times10^{12}\:$G; \citet{hui06b} prefer a two black body fit with temperatures and effective radii T$_1$ = 2.6$\times$10$^6$ K, T$_2$ = 5.0$\times$10$^6$ K, R$_1$ = 3.3 km, and R$_2$ = 0.75 km.  \citet{hui06b} place an upper limit of 2.9$\times$10$^{-13}$ ergs cm$^{-2}$ s$^{-1}$ on the 
0.5 - 10.0 keV X-ray flux from an associated wind nebula.   An extremely stringent constraint has been placed on the radio luminosity of a wind nebula,  three orders of magnitude below what would be expected if the stellar remnant were an energetic young pulsar \citep{gaensler00}.  The resulting stringent radio limit on ${\dot{E}}$ suggests a high magnetic field ($B > 6.4\times 10^{13}\,$ G), which in turn invites comparison of this object with AXPs and SGRs.  Such a comparison is premature, however, given the different spectral and temporal properties of RX~J0822--4300.  Velocity is not a discriminator, as AXPs and SGRs show a large range of inferred transverse velocities, from
$<100 \kms$\ to $\sim 2900 (3\, {\rm kyr}/t) \kms$ \citep{gaensler00b}.


The object most similar to RX~J0822--4300 is the central stellar remnant in Cas A\@.  Cas A is thought to be the result of the core-collapse explosion of a comparably massive star \citep{willingale03}, with oxygen its most abundant nucleosynthesis product.  Its stellar remnant has similar spectral properties to the one in Puppis A, though the one in Cas A is slightly hotter, as should be the case for a neutron star one-tenth the age \citep{pavlov02}.  It too shows no evidence for either a wind nebula or strong temporal variability.\footnote{As for RX~J0822--4300, the claimed periodicity of 12 ms of the Cas A source has not been confirmed \citep{murray02}.}  Its transverse velocity is inferred to be $330 \kms$\ for a distance of 3.4 kpc \citep{thorstensen01}, perpendicular to the most pronounced asymmetry axis in the remnant, defined by the ``jet" and ``counter jet" \citep{hwang04}.

The most challenging problem is to explain how neutron stars can 
have recoil velocities as high as we have measured for RX~J0822--4300. 
Numerous mechanisms have been proposed to provide kicks to nascent 
neutron stars during a supernova explosion.  Generally speaking, these 
fall into three broad categories:  electromagnetically driven, 
neutrino/magnetic-field driven, and hydrodynamically driven. A 
review of the physics of each can be found in \citet{lai01}. 
Briefly, in the electromagnetically driven mechanism, radiation from 
an off-centered rotating magnetic dipole imparts a gradual 
acceleration to the neutron star along its spin axis. In the 
neutrino/magnetic-field driven mechanism, the kick is produced by 
asymmetric neutrino emission arising in a strong magnetic field.  The 
hydrodynamically driven mechanism relies on asymmetric matter 
ejection resulting from hydrodynamic instabilities during the 
explosion.  The first two mechanisms lead naturally to alignment of 
the kick direction close to the neutron star spin 
axis.  Electromagnetically driven kicks can generate a 
spin-aligned kick of 1,000 km s$^{-1}$, but require millisecond spin 
periods do to so.  Neutrino/magnetic-field driven kick models do not produce velocities 
higher than $\sim 250  \kms$\ and require magnetic fields in excess 
of 10$^{15}$ G\@.    Both these mechanisms seem unlikely candidates 
for the origin of RX~J0822--4300.

Hydrodynamically driven shocks can in principle 
also provide kick velocities in excess of 1,000 km s$^{-1}$ and do 
not require spin-kick alignment.  \citep{burrows96,  burrows06, 
burrows07, scheck04, scheck06}.
However, a spin-kick alignment can occur during a hydrodynamically 
driven kick if rotation is dynamically important for the core collapse 
and explosion (which in turn requires the initial spin period to be 
less than 1 ms).  Alternatively, alignment can occur hydrodynamically 
from rotational averaging of the transverse momentum from small 
thrusts, provided the kick duration is substantially longer than the 
rotation period of the proto-neutron star \citep{spruit98}. 
Spin-kick alignment or near-alignment seems to be a common trait of 
isolated pulsars and neutron stars associated with pulsar wind 
nebulae \citep{ng04, johnston05, wang06}, but appears to be less 
common in neutron stars in binary systems \citep{wang06}.

Evidence regarding spin-kick alignment of RX~J0822--4300 is mixed.  The 
strongest supporting evidence is the existence of
of a bipolar cavity observed in \hi\ centered on the star and oriented approximately 
along the direction of motion \citep{reynoso03}.  These are postulated to arise from
oppositely directed jets; they have swept out approximately 2 M$_{\odot}$ of material.
The absence of readily detectable pulsations argues against alignment:   
if the system were aligned, the apparent space velocity vector, 
largely in the plane of the sky, provides an ideal viewing geometry 
for detecting pulsations.  Alternatively, the marginal detection requires alignment not only of the spin
and kick directions, but also the magnetic field vector.  

Independent of the spin-kick alignment question, the high transverse velocity of 
RX~J0822--4300 together with its other properties challenge explosion models that
produce neutron-star kicks.
The high magnetic field required for the 
electromagnetically-driven mechanism is consistent with the absence 
of a wind nebula,  but the absence of a period in the 6--12 s 
range typical of magnetars argues against such a mechanism.  Electromagnetically
driven kicks are  generally viewed as unlikely because these require very fast (ms) spin
periods, which are not associated with isolated pulsars.
Kick mechanisms 
driven by neutrinos may be ruled out by the momentum balance between 
RX~J0822--4300 and the oxygen knots as well as by the high velocity. 
The presence of these knots, along with the lack of pulsations and 
the analogy with Cas A, all favor a non-aligned kick of hydrodynamic 
origin.  

Recent SN explosion modeling by \citet{burrows07} makes 
two important predictions.  This model posits a mechanism for 
core-collapse SNe relying on acoustic power generated in the inner core. 
The recoil mechanism is hydrodynamical, due either to acoustic power 
or an asymmetric neutrino flux.  First, their model provides an 
estimate of the  kick velocity: 
$v_k \sim 1000 \: E/(10^{51}\; {\rm ergs})\; {\rm sin}\, \alpha \kms$, 
where {\it E} represents the explosion energy, 
and sin$\, \alpha$\ parameterizes the anisotropy of the explosion. 
This expression predicts a correlation between explosion energy and 
neutron star recoil velocity.  Even for the most extreme anisotropy, 
sin$\, \alpha\sim 1$, this model suggests that the explosion 
producing Puppis A must have been more energetic than the canonical 
10$^{51}$ ergs, and could have been considerably more so.  It is 
currently thought that the explosion producing Cas A had energy 
2--4\,$\times 10^{51}$ ergs \citep{laming06}; thus by analogy, a similar 
explosion energy might be expected for Puppis A\@.  A higher explosion 
energy offers the benefit (for this model) of reducing the energy fraction 
that must be channeled into the 
kick.

A second prediction of the \citet{burrows07} model is that the proto-neutron star and a 
compensating amount of inner ejecta are expelled in opposite 
directions.  The inner ejecta in core-collapse SNe are thought to be 
predominantly iron.  The oxygen knots originate in an outer, hydrostatic burning layer, and 
move much more slowly than the initial velocity of material ejected 
from nearer the core.  There is no evidence for excess Fe opposite the 
stellar direction of motion, however, or anywhere else in Puppis A. 
If fast Fe ejecta are present towards the east-northeast (up an ambient
density gradient), one would expect these to become visible 
through interaction 
with the medium.  The only apparent ejecta concentration is a region 
of enhanced Si to the north-northeast of RX~J0822--4300, and not in 
line with its motion \citep{hwang07}.  If the O knots are the only 
manifestation of an explosion asymmetry, it is not apparent how the 
momentum of the inner ejecta could be efficiently transferred to 
them.  

\section{Conclusions}
As a summary, we may highlight the following points:
(1) A precise position for RX~J0822--4300, the presumed neutron star,
has been measured from three independent \chandra\ HRC observations
spread over a baseline of over five years.  
Comparison  among these yields a proper motion of 
$0\farcs 165 \pm 0\farcs 025\; {\rm yr}^{-1}$\ at  position angle $248\degr \pm \, 14\degr$.  
The uncertainty is dominated by the precision of position 
fits for two fiducial stellar X-ray sources used in the analysis.

(2) This motion implies a transverse velocity of 
$1570\pm 240\,(d/2\; {\rm kpc}) \kms$\ toward the west-southwest.  This direction is roughly opposite
to the motion of a handful of oxygen-rich optical filaments---presumably near-undiluted ejecta 
from the outer core of the progenitor---that are scattered throughout the northeast 
quadrant of Puppis A\@.  Furthermore, backwards extrapolation of the motion of RX~J0822-4300
indicates an origin consistent with the measured expansion center for the O-rich
filaments.    This completes---in at least
one SNR---what is becoming almost a  canonical picture for
core-collapse supernovae:  an asymmetric supernova explosion accompanied by recoil of the compact stellar remnant, here almost surely a neutron star.   

(3) The kinetic energy associated with the transverse motion of RX~J0822-4300
is $\sim 3\times 10^{49}\; (d/2\; {\rm kpc})^2\; {\rm ergs}$, only about 3\% \ of the total 
of $\sim 10^{51} \; {\rm ergs}$\ expected in a typical supernova.   Some 2--3 dozen O-rich knots
like those now glowing optically are sufficient to balance the momentum of the neutron star.

(4) The physics of the explosion mechanism necessary to produce such a fast neutron star remains elusive, but the high kick velocity and lack of apparent pulsations from RX~J0822-4300 do constrain possible models.   Both these observations argue against  neutrino/magnetic-field driven 
or electromagentically driven mechanisms for the kick.  
The most likely candidate appears to be some mechanism through 
which hydrodynamic instabilities in the explosion lead to recoil of the compact remnant.
However, the most specific such model, that of \citet{burrows07}, is strained to explain both the 
high kick velocity and the apparent absence of iron-rich ejecta from the inner core of the 
Puppis A progenitor.

A complete kinematic study  of Puppis A, including both the oxygen knots and the stellar remnant, 
will be interesting and should be carried out.   CCD data should enable measurement of 
the motions for significantly more oxygen knots than the handful used by \citet{winkler88},
who based their study on only 11 individual knots whose motions could be measured on
photographic plates.   A third-epoch \chandra\ observation with the HRC-I and/or HRC-S 
would further cement the kinematic picture.

\acknowledgments
We have benefitted from extensive correspondence with Ralph Kraft and Almus Kenter 
of the \chandra\  X-ray Center at SAO regarding the \chandra\ HRC instrument, position fitting, and systematic effects.  
We also gratefully acknowledge valuable discussions of some statistical and computational issues 
attending this problem with John Emerson and Daniel Scharstein of Middlebury College, and
suggestions from an anonymous referee that helped clarify our presentation. 
This research has been supported through the {\it Chandra} program by NASA grant GO4-5062X,
with additional support from NSF grant AST-0307613.
 

\clearpage

\begin{deluxetable}{ccllc}
\tablecaption{{\it Chandra} HRC Observations of RXJ0822--4300. }
\tablewidth{0pt}
\tablehead{
\colhead{Obs ID} &\colhead{Sequence No} & \colhead{Detector} &\colhead{Date}  & \colhead{Exposure (ks)}
}

\startdata
\phn749  & 500045 & HRC-I & 1999 Dec 21-22& 15.9  \\
1851 & 500122 & HRC-S  & 2001 Jan 25 &19.5   \\
4612 & 500437 & HRC-I  &  2005 Apr 25 & 33.1  \\
\enddata


\end{deluxetable}

\begin{deluxetable}{cccllcccccccc}
\tabletypesize{\footnotesize}
\rotate
\tablewidth{0pc}
\tablecaption{Astrometric Reference Stars from UCAC2 Catalog\tablenotemark{a}}

\tablehead{
\multicolumn{2}{c}{Designation} &
\colhead{\phd} &
\multicolumn{2}{c}{Position} &
\colhead{\phd} &
\multicolumn{2}{c}{Pos'n Uncertainty} &
\colhead{\phd} &
\multicolumn{2}{c}{Proper Motion} & 
\colhead{\phd} &
\colhead{Distance}  \\

\cline{1-2} \cline{4-5} \cline{7-8} \cline{10-11}

\colhead{Short} & 
\colhead{2UCAC No} &
\colhead{} &
\colhead {RA (2000.)} &
\colhead {Dec (2000.)} &
\colhead{} &
\colhead {$\sigma_{\rm RA}$} & 
\colhead{ $\sigma_{\rm Dec}$} &
\colhead{} &
\colhead {$\mu_{\rm RA}$} & 
\colhead {$\mu_{\rm Dec}$} & 
\colhead{} &
\colhead{from NS\tablenotemark{b}} \\

\colhead{} & 
\colhead{} &
\colhead{} &
\colhead {} &
\colhead {} &
\colhead{} &
\colhead{(mas)} & 
\colhead{(mas)} &
\colhead{} &
\colhead {(${\rm mas\; yr^{-1}})$} & 
\colhead {(${\rm mas\; yr^{-1}})$} & 
\colhead{} &
\colhead{($\arcmin)$} 
}
\tablenotetext{a}{\citet{zacharias03}}
\tablenotetext{b}{Angular separation from the central neutron star RXJ0822-4300.}

\startdata

A & 13302738 &  & $8^{\rm h}21^{\rm m}46\fs294 $ & 
$-43\arcdeg 02\arcmin 03\farcs 64$   & & 15 & 15 & & $-16.0\pm5.2 $ &
$-1.7\pm 5.2 $ & & 2.7 \\

C & 13302743 & & $8\phm{^{\rm h}}21\phm{^{\rm m}}48.875 $ & 
$-43\phm{\arcdeg} 01\phm{\arcmin} 28. 34$ & & 15 & 15 & & $-65.5\pm4.7 $ &
$-7.0\pm 4.7 $ & & 2.0 \\
   
B &  13520024 & & $8\phm{^{\rm h}}22\phm{^{\rm m}} 24.004 $ & 
$-42\phm{\arcdeg} 57\phm{\arcmin} 59.36$  & &24 & 15 & &  $-\phn4.2\pm5.2 $ &
$14.8\pm5.2 $ & & 5.4 \\

\enddata

\end{deluxetable}

\begin{deluxetable}{cccccccrccc}
\tabletypesize{\footnotesize}
\rotate
\tablewidth{0pc}
\tablecaption{Properties of \chandra\ X-ray Sources and Positions for Optical Counterparts}

\tablehead{
\colhead{} & 
\colhead{} &
\colhead{} &
\colhead{\phd} &
\multicolumn{2}{c}{X-ray Position\tablenotemark{a,b}} &
\colhead{\phd} &
\colhead{Count Rate} &
\colhead{\phd} &
\multicolumn{2}{c}{UCAC2 Position\tablenotemark{c}} \\

\cline{5-6}  \cline{10-11}

\colhead{Obs ID} & 
\colhead{Epoch} &
\colhead{Source} &
\colhead{} &
\colhead {RA (2000.)} &
\colhead {Dec (2000.)} &
\colhead{} &
\colhead {counts ks$^{-1}$} & 
\colhead{} &
\colhead {RA (2000.)} &
\colhead {Dec (2000.)} 

}
\tablenotetext{a}{Position determined from fit of appropriate PSFs to HRC data (see text).}
\tablenotetext{b}{Uncertainty ($1\sigma$) in last two digits in parentheses.}
\tablenotetext{c}{\citet{zacharias03}; corrected for proper motion.}

\startdata

\phn 749 & 1999.98 & NS &  & $8^{\rm h}21^{\rm m}57\fs410(01) $ & 
$-43\arcdeg 00\arcmin 16\farcs 66(01)$   & & 195.5\phn\phn\phn\phn & &  & \\
   
  &  & A &  & $8\phm{^{\rm h}}21\phm{^{\rm m}}46.284(10) $ & 
$-43\phm{\arcdeg} 02\phm{\arcmin} 03.13(24)$ & & 3.0\phn\phn\phn\phn & & $8^{\rm h}21^{\rm m}46\fs294 $ &
$-43\arcdeg 02\arcmin 03\farcs 64$ \\
   
  &  & B &  & $8\phm{^{\rm h}}22\phm{^{\rm m}} 23.915(24) $ & 
$-42\phm{\arcdeg} 57\phm{\arcmin} 59.48(21)$  & & 6.7\phn\phn\phn\phn & 
& $8\phm{^{\rm h}}22\phm{^{\rm m}}24.004 $ &
$-42\phm{\arcdeg} 57\phm{\arcmin} 59.36$

\vspace{0.1in}\\

1851  &  2001.07 & NS &  & $8\phm{^{\rm h}}21\phm{^{\rm m}} 57.393(01) $ & 
$-43\phm{\arcdeg} 00\phm{\arcmin} 16.68(01)$  & & 315.4\phn\phn\phn\phn & & & \\

  &  & A &  & $8\phm{^{\rm h}}21\phm{^{\rm m}} 46.331(06) $ & 
$-43\phm{\arcdeg} 02\phm{\arcmin} 03.41(08)$  & & 4.9\phn\phn\phn\phn & 
& $8\phm{^{\rm h}}21\phm{^{\rm m}}46.292 $ &
$-43\phm{\arcdeg} 02\phm{\arcmin} 03.64$ \\

  &  & B &  & $8\phm{^{\rm h}}22\phm{^{\rm m}} 23.871(13) $ & 
$-42\phm{\arcdeg} 57\phm{\arcmin} 59.04(13)$  & & 10.0\phn\phn\phn\phn & 
& $8\phm{^{\rm h}}22\phm{^{\rm m}}24.003 $ &
$-42\phm{\arcdeg} 57\phm{\arcmin} 59.34$

\vspace{0.1in}\\

4612  &  2005.32 & NS &  & $8\phm{^{\rm h}}21\phm{^{\rm m}} 57.364(01) $ & 
$-43\phm{\arcdeg} 00\phm{\arcmin} 16.99(01)$  & & 209.0\phn\phn\phn\phn & & & \\

  &  & A &  & $8\phm{^{\rm h}}21\phm{^{\rm m}} 46.296(06) $ & 
$-43\phm{\arcdeg} 02\phm{\arcmin} 03.45(04)$  & & 3.5\phn\phn\phn\phn & 
& $8\phm{^{\rm h}}21\phm{^{\rm m}}46.286 $ &
$-43\phm{\arcdeg} 02\phm{\arcmin} 03.65$ \\

  &  & B &  & $8\phm{^{\rm h}}22\phm{^{\rm m}} 24.005(09) $ & 
$-42\phm{\arcdeg} 57\phm{\arcmin} 59.12(28)$  & & 5.4\phn\phn\phn\phn & 
& $8\phm{^{\rm h}}22\phm{^{\rm m}}24.002 $ &
$-42\phm{\arcdeg} 57\phm{\arcmin} 59.28$ \\

\enddata

\end{deluxetable}

\begin{deluxetable}{ccccccc}
\tablecaption{Position and Proper Motion Measurements for RXJ0822-4300.\tablenotemark{a} \label{tbl-2}}
\tablewidth{0pt}
\tablehead{
\colhead{} & \colhead{Baseline}   & 
\colhead{} &   \colhead{Displacement} & 
\colhead{Proper Motion} &\colhead{} \\

\colhead{Obs IDs} & \colhead{(yr)} & \colhead{Method} & \colhead{(mas)} & \colhead{$({\rm mas\; yr}^{-1})$} &
\colhead{Position Angle\tablenotemark{b}}  
}

\startdata
\phn749--4612 & 5.34 & shift only\tablenotemark{c} & $752\pm 131$ & $141 \pm 24$ & 
$250\degr \pm 14\degr$ \\

\phn749--4612 & 5.34 & rot-scale\tablenotemark{d} & $990\pm 149$ & $185 \pm 28$ & 
$254\degr \pm 16\degr$ \\

1851--4612 & 4.25 & rot-scale\tablenotemark{d} & $716\pm 100$ & $168 \pm 23$ & 
$240\degr \pm 13\degr $ \\

Mean\tablenotemark{e} & & & & $165\pm  25 $ & $248\degr \pm  14\degr $\\

\enddata
\tablenotetext{a}{Corrected position for RX~J0822--4300 in 2005 April observation:  
RA(2000.)$\, = 8^{\rm h}21^{\rm m}57\fs355$,   Dec(2000.)$\, = -43\arcdeg 00\arcmin 17\farcs 17$.}
\tablenotetext{b}{Measured north through east.}
\tablenotetext{c}{Small RA and Dec shifts alone are used to align stars A and B in the two images (\S3.2).}
\tablenotetext{d}{Images at all three epochs are placed on a common absolute frame through a 
linear transformation that allows rotation and a uniform scale change in addition to 
translations to align stars A and B with their optical positions (\S3.3).}
\tablenotetext{e}{Mean is the unweighted average of all three measurements.}
\end{deluxetable}


\clearpage






\begin{figure}
\plotone{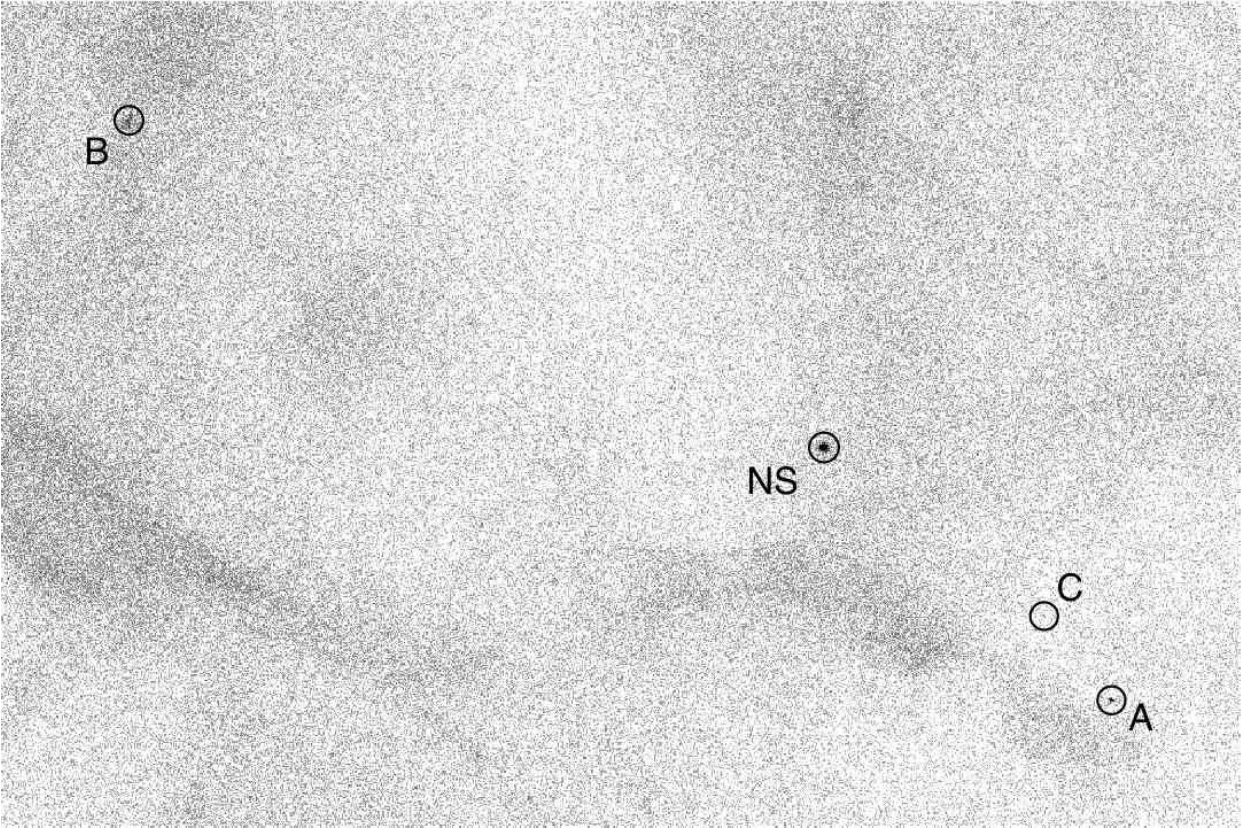}
\caption{ The 2005 epoch {\it Chandra} HRC-I image with sources marked by 12\arcsec\  diameter circles.
NS is the presumed neutron star, A and B correspond to fiducial stars used in our analysis, 
and C marks the position of a third star detected at only the $\sim 2\sigma$ level 
in 2005 (see also Fig.\ 2).  The field measures 
$8\farcm 6 \times 5\farcm 7$\ and is oriented north up, east left.}
\end{figure}

\begin{figure}
\plotone{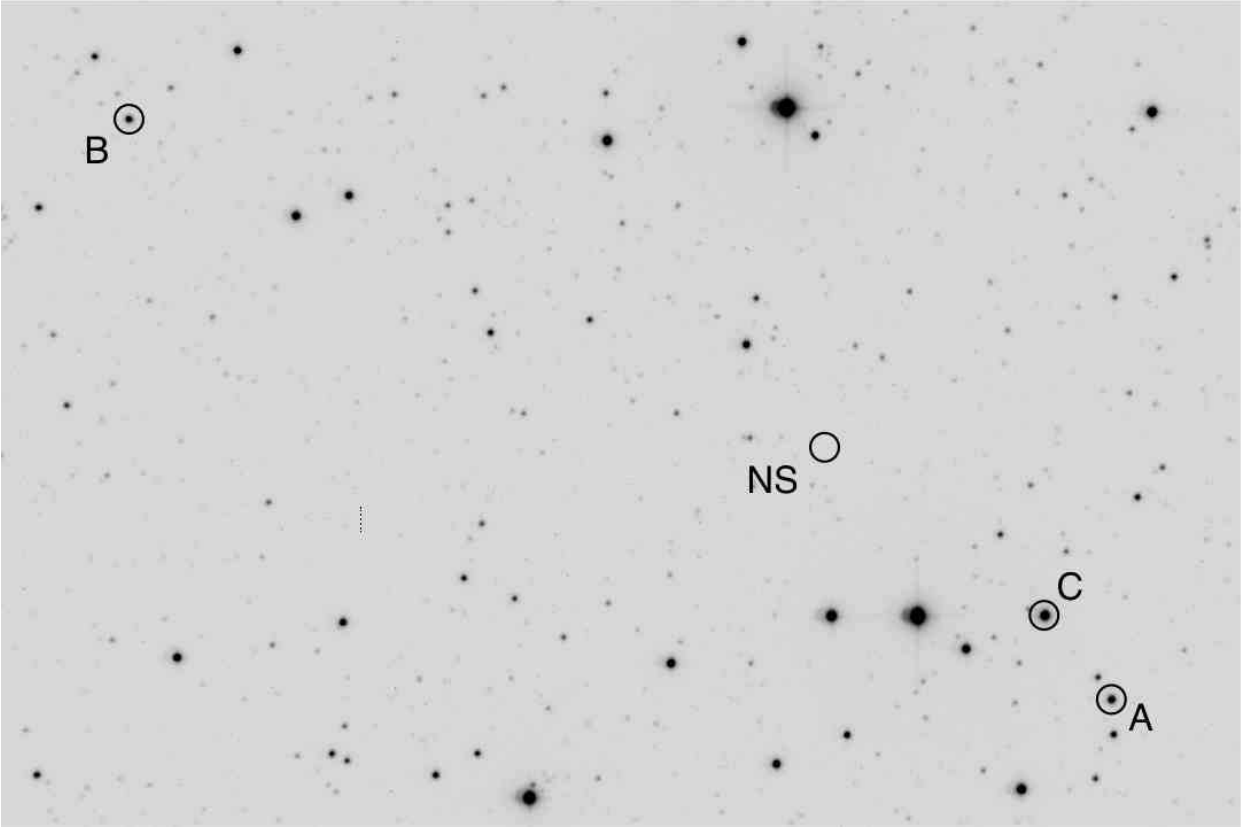}
\caption{ Optical image of the identical field as Figs.\ 1 and 4, with the astrometric
reference stars marked (the same overlay as Fig.\ 1).   The image was obtained at the 
CTIO 0.9 m telescope using a narrow-band red continuum filter.}
\end{figure}

\begin{figure}
\epsscale{0.75}
\plotone{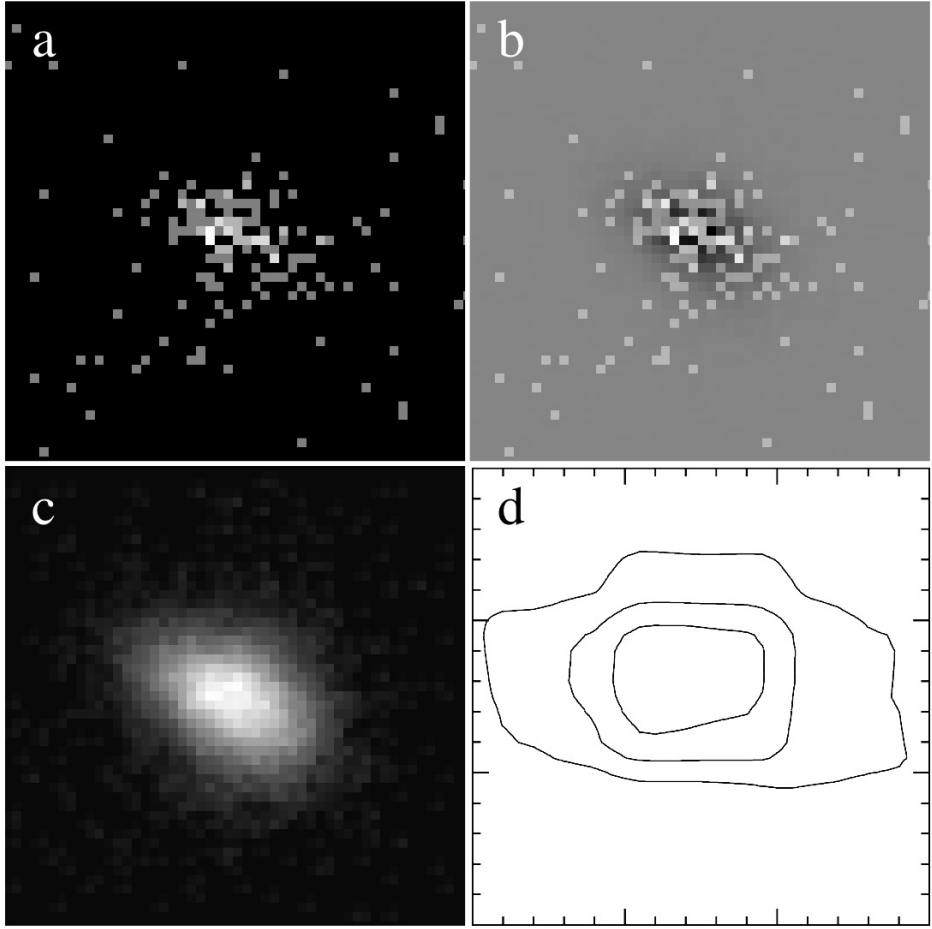}
\epsscale{1.0}
\caption{Example of a typical source position fit---this for star A at epoch 2005 
(Obs ID 4612):\ ({\it a}) The HRC-I data displayed at full resolution; 
the small squares are individual HRC pixels, 
0\farcs 1318 square, and the field shown here (and in panels {\it b} and {\it c}) is 50 pixels = 6\farcs 6 square.
({\it c})  The PSF as simulated using ChaRT and MARX for a source with the same off-axis angle and
azimuth as star A in this observation.   ({\it b}) The difference between the actual data (panel {\it a})
and the best-fit source model after convolution with the PSF in panel {\it c}.  The position, intensity, 
and constant background level have been adjusted in the fit.    ({\it d}) A map of confidence contours for the 
position fit; the major ticks represent 1 HRC pixel, and the full field here is only 3 pixels =  0\farcs 40
square.  The Cash statistic has been computed at each ({\it x, y}) point in a fine grid, with the 
intensity and background readjusted to give the minimum value at that point.  The contours 
correspond to 1, 2, and $3\,\sigma$\ (increments of 2.3, 6.2, and 11.8 above the minimum statistic value).  
The contours are not circular, but the $1\,\sigma$ uncertainty in any direction is $\lesssim \pm 0.5\; {\rm pixels} \approx 0\farcs 065$\@.  Similar fits were carried out for the NS and for stars A and B at
all three epochs, to obtain the values given in Table 3.} 
\end{figure}

\begin{figure}
\plotone{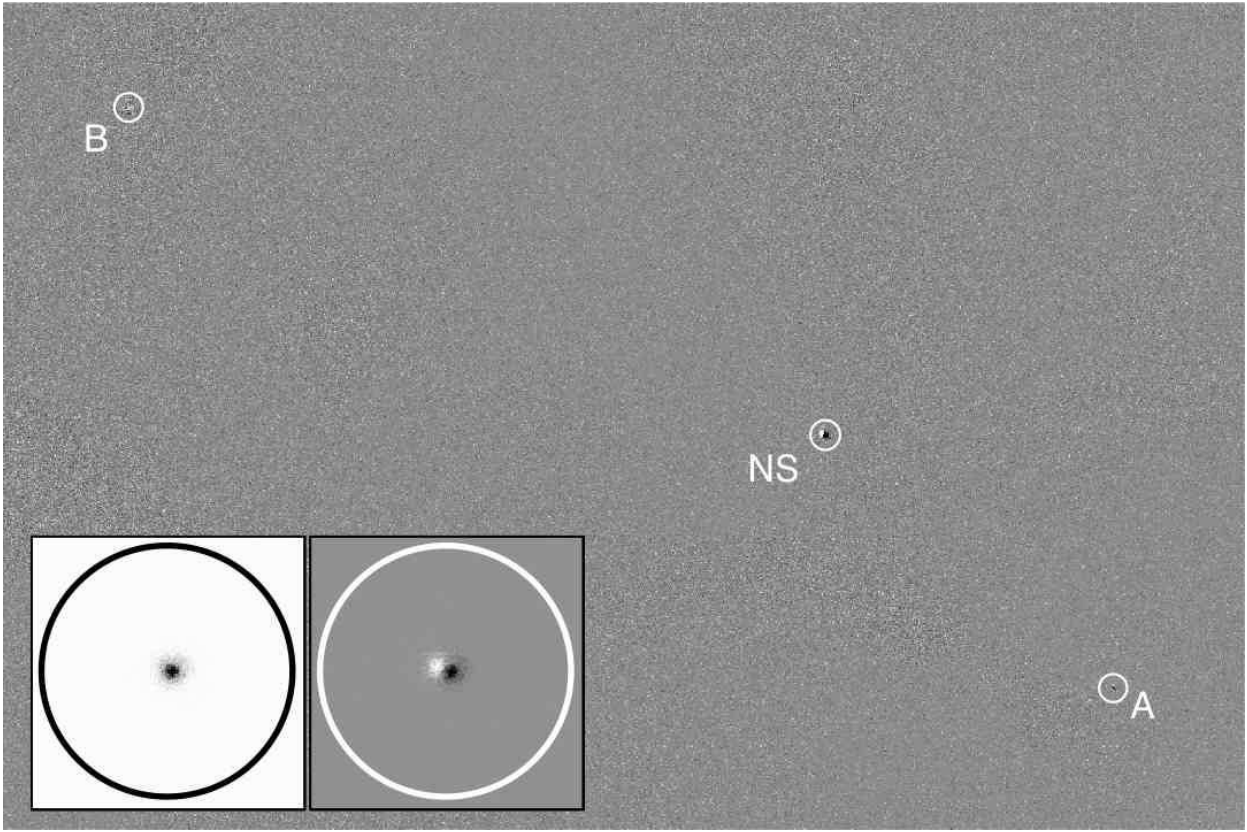}
\caption{ Difference between the 1999 epoch (events white) and 2005 epoch 
(events black) HRC-I images, registered and scaled by the exposure time.  The field is the
same as in Figs. 1 and 2.
The reference stars A and B  largely disappear in the difference image, 
but RX~J0822--4300 shows a noticeable displacement between the two epochs.  
This motion is more evident in the detailed view shown in the inset: ({\it left})  the 2005 epoch 
image alone, and ({\it right}) and the difference image.  The
images are the same as those shown in Figs..\ 1 and 4, but the stretch has been changed
to emphasize the narrow ($\sim 0\farcs 5$) point-spread function.  The circles in the inset are 12\arcsec\ in diameter, the same as in the full image and in Fig. 1.
Proper motion toward the west-southwest is apparent, with a displacement
roughly twice the width of the PSF.}
\end{figure}

\begin{figure}
\plotone{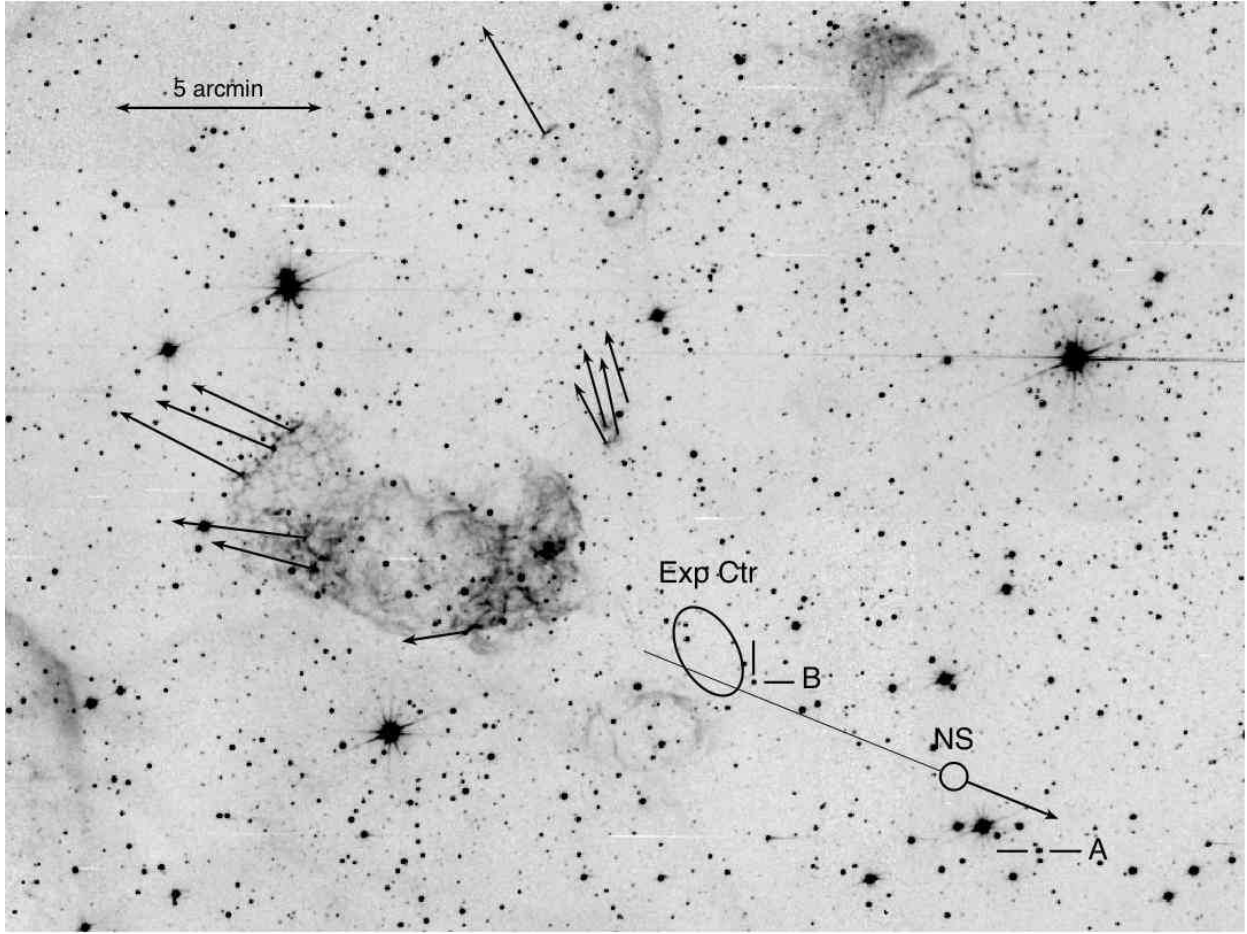}
\caption{Optical image of the central region of Puppis A in \oiii\lam 5007,  
showing proper motions of the O-rich filaments and RX~J0822--4300.  
The arrows indicate proper-motion vectors for $\sim 1000$\ yr, 
and the ellipse shows the 90\%-confidence contour for the expansion 
center.  \citep[This is a wider-field version of Fig.~1 from][overlaid on a more recent CCD image.]{winkler88}
The circle marked NS shows the present position for the presumed neutron star, 
RX~J0822--4300, and the attached vector indicates its motion over 1000 yr at the rate we have
measured.   Backwards extrapolation of its present motion (lighter line) passes well within the 
error ellipse for the expansion center  of the O-rich filaments.}  
\end{figure}

\clearpage

\end{document}